\RequirePackage{pdf14}
\documentclass{ifacconf}
\usepackage{xcolor}
\usepackage{graphicx}      
\usepackage{natbib}        
\usepackage{amssymb}
\usepackage{amsmath}
\usepackage{commath}
\usepackage{cuted}
\usepackage{balance}
\usepackage{multicol}
\graphicspath{{figures/}}

\newcommand{\vc}[1]{\mathrm{vec}(#1)}

\begin{document}
\begin{frontmatter}

\title{Finite Dimensional Koopman Form of Polynomial Nonlinear Systems\thanksref{footnoteinfo}} 

\thanks[footnoteinfo]{This work has received funding from the European Research Council (ERC) under the European Union’s Horizon 2020 research and innovation programme (grant agreement nr. 714663) and from the European Union within the framework of the National Laboratory for Autonomous Systems (RRF-2.3.1-21-2022-00002).}

\author[First]{Lucian C. Iacob} 
\author[First]{Maarten Schoukens} 
\author[First,Second]{Roland T\'oth}

\address[First]{Control Systems Group, Eindhoven University of Technology, Eindhoven, The Netherlands \\e-mail: l.c.iacob@tue.nl, m.schoukens@tue.nl, r.toth@tue.nl}
\address[Second]{Systems and Control Laboratory, Institute for Computer Science and Control, Budapest, Hungary}

\begin{abstract}                
The Koopman framework is a popular approach to transform a finite dimensional nonlinear system into an infinite dimensional, but linear model through a lifting process, using so-called observable functions. While there is an extensive theory on infinite dimensional representations in the operator sense, there are few constructive results on how to select the observables to realize them. When it comes to the possibility of finite Koopman representations, which are highly important form a practical point of view, there is no constructive theory. Hence, in practice, often a data-based method and ad-hoc choice of the observable functions is used. When truncating to a finite number of basis, there is also no clear indication of the introduced approximation error. In this paper, we propose a systematic method to compute the finite dimensional Koopman embedding of a specific class of polynomial nonlinear systems in continuous-time such that, the embedding, without approximation,  can fully represent the dynamics of the nonlinear system.
\end{abstract}


\end{frontmatter}

\section{Introduction}
In most engineering fields, due to increasing performance demands, tackling the nonlinear behaviour becomes more and more important. However, the available methods in the field of nonlinear control (e.g. feedback linearization, backstepping, sliding mode
control \citep{Khalil:02}) are generally complex to design, only offer stability guarantees, and performance shaping of the closed-loop has yet to be achieved. This is in contrast to the systematic and powerful tools available for \textit{linear time invariant} (LTI) systems. However, using LTI control tools on linearized models offers limited performance when the system evolves away from the operating region. Hence, there is an increasing need to extend the powerful LTI control design and modelling framework to address nonlinear systems. As such, there is a significant interest in finding globally linear surrogate models of nonlinear systems.
\par One of the more promising approaches to achieve this is given by the Koopman framework \citep{Brunton:21}, \citep{Bevanda:21}, \citep{Mauroy:20}, where the concept is to project the original nonlinear state space  representation to a higher dimensional (possibly infinite) but linear space, through observable functions. The Koopman operator is a linear operator and governs the dynamics of the observables. The Koopman framework shows promising results in its application to real-world analysis and control applications (e.g. mechatronic systems \citep{Abraham:19}, \citep{Cisneros:20}, distributed parameter systems \citep{Klus:20}). For practical use, a finite number of observables needs to be selected. These are then used to construct time shifted data matrices, to compute via least-squares the matrix representation of the Koopman operator. This technique is known as \textit{extended dynamic mode decomposition} (EDMD) \citep{Williams:15}. However, the main problem is that the choice of the observables is heuristic and there are no guarantees on the quality of the resulting model. To tackle this, one solution is to use data-driven techniques to learn the lifting from data, in order to circumvent the manual selection of observables \citep{Lusch:18}, \citep{Iacob:21}. Nevertheless, this is still an approximation and the questions on how to embed the nonlinear system into an exact linear finite dimensional lifted representation and when this is possible at all are still open. This is an important aspect, because, for control purposes, having an exact finite dimensional embedding allows for the application of the available control tools for linear systems. Moreover, if there exist approximation errors in the model that cannot be quantified, the expected performance will not be achieved. To tackle this, there have been attempts to connect the Koopman framework to immersion \citep{Wang:20} and Carleman linearization, in order to obtain a clear way of computing the observables. However, in the immersion approach, the existence of a finite dimensional fully linear lifting depends heavily on the observability property of the system and, in general, the resulted embedding contains a nonlinear output injection \citep{Krener:83}, \citep{Jouan:03}. For the Carleman linearization \citep{Kowalski:91}, while it offers a systematic way of computing the lifting functions, the resulting embedding is still an infinite dimensional model that needs to be trimmed. \par 
The present paper discusses a novel method to systematically convert a polynomial nonlinear system to an exact finite dimensional linear embedding. Starting from the idea of the simple 2-dimensional example shown in \citep{Brunton:21}, we introduce a state-space model where the state equation is described by a lower triangular polynomial form. We prove that there always exists an exact finite dimensional Koopman representation and we show how to systematically compute it. Furthermore, we also show that, once the autonomous part of the nonlinear system is fully embedded, the extension to systems with inputs is trivial and can be performed in a separate step. Using an example system, we demonstrate that the lifted Koopman model can fully capture the original dynamics, both in an autonomous operation and in the presence of inputs. \par 
The paper is structured as follows. Section \ref{sec:embedding} describes the Koopman framework and details the proof and steps needed to obtain the finite embedding. In Section \ref{sec:example}, we discuss the example and showcase the simulation results. In Section \ref{sec:conclusion}, conclusions on the presented results are given together with outlooks on future research.
\section{Finite dimensional embedding}\label{sec:embedding}
The present section details the Koopman framework and showcases the proposed method to compute an exact finite dimensional embedding. Additionally, we discuss the extension to systems with inputs.
\subsection{Koopman framework}
Consider the autonomous nonlinear system:
\begin{equation}\label{eq:nl_aut_general}
\dot{x}=f(x),
\end{equation}
with $x:=x(t)$ denoting the state, $t\in\mathbb{R}$ represents the time and $f:\mathbb{R}^{n_\mathrm{x}}\rightarrow\mathbb{R}^{n_\mathrm{x}}$ is the nonlinear vector field which we consider to be a Lipshitz continuous function. Given an initial condition $x(0)\in\mathbb{X}\subseteq\mathbb{R}^{n_\mathrm{x}}$, the solution $x(t)$ can be described as:
\begin{equation}
x(t)=F(t,x(0)):=x(0)+\int^t_0 f(x(\tau))\dif \tau.
\end{equation}
It is assumed that $\mathbb{X}$ is compact and forward invariant under the flow $F(t,\cdot)$, such that $F(t,\mathbb{X})\subseteq \mathbb{X}, \forall t\geq 0$. Introduce the family of Koopman operators $\{\mathcal{K}^t \}_{t\geq 0}$ associated to the flow $F(t,\cdot)$ as:
\begin{equation}
\mathcal{K}^t\phi(x(0))=\phi\circ F(t,x(0)),\quad \phi\in \mathcal{F},
\end{equation} 
where, $\mathcal{F}\subseteq\mathcal{C}^1$ is a Banach function space of continuously diferetiable functions and $\phi:\mathbb{X}\rightarrow\mathbb{R}$ is a scalar observable function. As the flow $F$ is uniformly Lipshitz and $\mathbb{X}$ a compact forward-invariant set, the Koopman semigroup $\{\mathcal{K}^t \}_{t\geq 0}$ is strongly continuous on $\mathcal{F}$ \citep{Mauroy:20}. Thus, we can describe the infinitesimal generator $\mathcal{L}:\mathcal{D}_{\mathcal{L}}\rightarrow\mathcal{F}$ associated to the Koopman semigroup of operators \citep{Lasota:94}, \citep{Mauroy:20} as:
\begin{equation}
\mathcal{L}\phi(x_0)=\lim_{t \downarrow 0} \frac{\mathcal{K}^t\phi(x(0))-\phi(x(0))}{t},\quad \phi\in\mathcal{D}_{\mathcal{L}},
\end{equation}
where $\mathcal{D}_{\mathcal{L}}$ is a dense set in $\mathcal{F}$. Note that, as described in \citep{Lasota:94}, the generator $\mathcal{L}$ is a linear operator. Through the infinitesimal generator we can thus describe the dynamics of observables as follows:
\begin{equation}\label{eq:observable_inf}
\dot{\phi}=\frac{\partial \phi}{\partial x}f=\mathcal{L}\phi,
\end{equation}
which is a linear infinite dimensional representation of the nonlinear system \eqref{eq:nl_aut_general}. If there exists a finite dimensional Koopman subspace $\mathcal{F}_{n_{\mathrm{f}}}\subseteq\mathcal{D}_{\mathcal{L}}$, such that the image of $\mathcal{L}$ is in $\mathcal{F}_{n_{\mathrm{f}}}$, then, given the set of lifting functions as basis of $\mathcal{F}_{n_{\mathrm{f}}}$, $\forall \phi\in\Phi$, $\mathcal{L}\phi\in\text{span}\{\Phi\}$. Thus, the following relation holds:
\begin{equation}
\dot{\phi}_j=\mathcal{L}\phi_j = \sum^{n_\mathrm{f}}_{i=1}L_{ij}\phi_i,
\end{equation}
where $L$ denotes the matrix representation of $\mathcal{L}$ and the coordinates of $\mathcal{L}\phi_j$ in the basis $\Phi$ are contained in the column $L_{\cdot j}$ Let $A=L^\top \in \mathbb{R}^{n_{\mathrm{f}}\times n_{\mathrm{f}}}$, and, based on \eqref{eq:observable_inf}, the lifted representation of \eqref{eq:nl_aut_general} is given by:
\begin{equation}
\dot{\Phi}(x)=\frac{\partial \Phi}{\partial x}(x)f(x)=A\Phi(x).
\end{equation}
Thus, one can formulate conditions for the existence of a finite dimensional embedding of \eqref{eq:nl_aut_general} as:
\begin{subequations}\label{eq:conditions}
\begin{align}
\dot{\Phi} &\in\text{span}\{\Phi\}, \label{eq:conditions_b}\\
\intertext{which is equivalent to} 
\frac{\partial \Phi}{\partial x}f&\in\text{span}\{\Phi\}. \label{eq:conditions_a} 
\end{align}
\end{subequations}
However, the major question is how to compute $\Phi$ such that the conditions \eqref{eq:conditions} are true. In the Koopman framework, to recover the original states of \eqref{eq:nl_aut_general}, the existence of a back transformation $\Phi^{\dagger}(\Phi(x))=x$ is often assumed. For simplicity, this is achieved by adding an extra condition to \eqref{eq:conditions}, namely that the original states are contained in $\Phi$, i.e., the identity function is part of $\Phi$. Next, in order to explicitly write the LTI dynamics given by the Koopman form, let $z(t) = \Phi(x(t))$. Then, an associated Koopman representation of \eqref{eq:nl_aut_general} is:
\begin{equation}\label{eq:koop_lti_gen}
\dot{z}=Az, \quad \text{with } z(0)=\Phi(x(0)).
\end{equation}
It is important to note that, by the existing theory, in general one cannot guarantee the existence of a finite dimensional Koopman invariant subspace $\mathcal{F}_{n_{\mathrm{f}}}$. In the sequel we show that, in case of systems described by a state-space representation where the state equation can be written in a lower triangular polynomial form, there always exists an exact finite dimensional Koopman  representation of the system in the form of \eqref{eq:koop_lti_gen} and this representation can be systematically computed.
\subsection{Exact finite embedding procedure}
Consider the nonlinear system \eqref{eq:nl_aut_general} to have the following structure:
\begin{equation}\label{eq:nl_aut}
\begin{split}
\dot{x}_1 &= a_1 x_1\\
\dot{x}_2 &= a_2 x_2 + f_2(x_1)\\
\dot{x}_3 &= a_3 x_3 + f_3(x_1,x_2)\\
&\vdots \\
\dot{x}_n &= a_n x_n + f_n(x_1,\dots,x_{n-1})
\end{split}
\end{equation}
where $f_n$ is given by:
\begin{equation}
f_n(x_1,\dots,x_{n-1})=\sum^{d_n}_{j_1=0}\dots\sum^{d_n}_{j_{n-1}=0}\alpha^n_{j_1\dots j_{n-1}}\prod^{n-1}_{i=1}x^{j_i}_i,
\end{equation}
with polynomial terms of the form $x^{j_1}_1\dots x^{j_{n-1}}_{n-1}$. It is assumed that the powers go up to $d_n$, for ease of derivation, but there is no restriction and each power can be arbitrarily large (but finite). It could be viewed that $d_n$ is the maximum power within the polynomial terms. Under these considerations, we can give the following theorem.
\begin{thm}\label{th1}
For an autonomous continuous-time nonlinear system that has a polynomial state-space representation in the form of \eqref{eq:nl_aut}, there exists an exact finite-dimensional lifting $\Phi:\mathbb{R}^{n_\mathrm{x}}\rightarrow\mathbb{R}^{n_\mathrm{f}}$, containing the states $x_i$, with $i\in\left\lbrace 1,\dots,n \right\rbrace$, such that \eqref{eq:conditions_b} holds true. 
\end{thm}
\begin{pf}
The theorem is proven by induction. First, we will consider the cases when $n=1,2,3,4$ and then we will show that if the statement of Theorem \ref{th1} holds for $n$-number of states then we can prove that it also holds for $n+1$.
\begin{itemize}
\item $n=1$ (first order system):
\begin{equation}\label{eq:nl_1state}
\dot{x}_1 = a_1 x_1
\end{equation}
Let $W_1 = \{x_1\}$ and $\Phi = \mathrm{vec}(W_1)$, i.e., $\Phi(x)=x_1$. It is trivial to see that condition \eqref{eq:conditions_b} holds true as $\dot{x}_1=a_1 x_1 \in \text{span}\{\Phi\}$.
\item $n=2$ (second order system): Notice that the dynamics defined by the $2^\mathrm{nd}$-order system are described by \eqref{eq:nl_1state}, together with 
\begin{equation}\label{eq:nl_2states}
\dot{x}_2 =a_2x_2 + \sum^{d_2}_{j_1=0}\alpha^2_{j_1}x^{j_1}_1.
\end{equation}
Here, superscript $2$ of the coefficient $\alpha^2_{j_1}$ denotes that it belongs to the $2^\mathrm{nd}$ state equation and not that the coefficient is raised to power 2. Let $V_2=\{x^{0}_1,\ldots, x^{d_2}_1\}$  and $W_2=\{x_2\} \cup V_2$, while $\Phi=\vc{W_1\cup W_2}$. By calculating $\dot{\Phi}$, we get the terms associated with $W_1$ and the terms  
\begin{equation} \label{eq:proof:1}
\frac{\dif}{\dif t} \left(x^{j_1}_1\right)=j_1x^{j_1-1}_1 \dot{x}_1=j_1a_1x^{j_1}_1
\end{equation}
originating from $V_2$.
It is easy to observe 
that all terms in \eqref{eq:proof:1} are already contained in $\Phi$ and $\dot{x}_2\in\mathrm{span}
\{\Phi\}$, hence condition \eqref{eq:conditions_b} holds true.
 \item $n=3$ (third order system): The dynamics of the $3^\mathrm{rd}$-order system are described by \eqref{eq:nl_1state}, \eqref{eq:nl_2states} and the following equation:
\begin{equation}\label{eq:nl_3states}
\dot{x}_3=a_3 x_3 + \sum^{d_3}_{j_1=0}\sum^{d_3}_{j_2=0}\alpha^3_{j_1,j_2}x^{j_1}_1 x^{j_2}_2.
\end{equation}
As performed previously, we take the nonlinear terms $x^{j_1}_1 x^{j_2}_2$ and add them to the set of lifting functions $V_3=\{x^{0}_1 x^{0}_2, \ldots, x^{d_3}_1 x^{d_3}_2\}$ and $W_3=\{x_3\} \cup V_3$, while $\Phi=\vc{W_1\cup W_2 \cup W_3}$. By calculating $\dot{\Phi}$, we get the terms associated with $W_1$, $W_2$ as before and 
\begin{align}
\frac{\dif}{\dif t}&\left(x^{j_1}_1 x^{j_2}_2\right)=j_1x^{j_1-1}_1 x^{j_2}_2 \dot{x}_1 + j_2x^{j_1}_1x^{j_2 -1}_2\dot{x}_2  \label{eq:proof:2}\\ 
&=(j_1a_1 + j_2a_2) \underbrace{x^{j_1}_1x^{j_2}_2}_{a}+j_2\sum^{d_2}_{\tilde{j}_1=0}\alpha^2_{\tilde{j}_1} \underbrace{x^{j_1+\tilde{j}_1}_1 x^{j_2 - 1}_2}_{b} \notag
\end{align}
originating from $V_3$. The following observations can be made:
\begin{itemize}
\item The terms $a$ are already contained in $V_3$. 
\item For the terms $b$, we can observe that the power $j_2$ decreases by 1 and $j_1$ increases by at most $d_2$. 
\end{itemize}
Introduce the operator $\mathfrak{D}_\mathrm{b}$ such that $\mathfrak{D}_\mathrm{b}(x^{j_1}_1 x^{j_2}_2)=\{x^{j_1+\tilde{j}_1}_1 x^{j_2 - 1}_2\}_{\tilde{j}_1=0}^{d_2}$, i.e., it gives the $b$ terms of \eqref{eq:proof:2}. Then let $V_3 \leftarrow V_3 \cup \mathfrak{D}_\mathrm{b}(V_3)$. Repeating the process, i.e., applying the time derivative again to $x^{j_1+\tilde{j}_1}_1 x^{j_2 -1}_2$ further decreases $j_2$ and increases the power of $x_1$, and at each step only terms of the form $a$ and $b$ are generated. Repeating the process for a finite number of steps gives that $\mathfrak{D}_\mathrm{b}(V_3) \setminus V_3
\subseteq\{x_1^{0},\ldots,x_1^{n_1}\}$. Hence, based on case $n=2$, we know that for $V_3 \leftarrow V_3 \cup \mathfrak{D}_\mathrm{b}(V_3)$ taking $W_3=\{x_3\} \cup V_3$ and $\Phi=\vc{W_1\cup W_2 \cup W_3}$ will ensure that  condition \eqref{eq:conditions_b} holds true.
\item  $n=4$ (fourth order system): The dynamics of the $4^\mathrm{th}$-order system is described by \eqref{eq:nl_1state}, \eqref{eq:nl_2states}, \eqref{eq:nl_3states}, together with:
\begin{equation}\label{eq:nl_4states}
\dot{x}_4=a_4 x_4+ \sum^{d_4}_{j_1=0}\sum^{d_4}_{j_2=0}\sum^{d_4}_{j_3=0}\alpha^4_{j_1, j_2, j_3}x^{j_1}_{1}x^{j_2}_2 x^{j_3}_3.
\end{equation}
To ease readability, let $\zeta_j = x^{j_1}_1 x^{j_2}_2$ with $j=j_1+(d_4+1)j_2+1$. This means that $j=1$ corresponds to $j_1=0,j_2=0$, $j=2$ corresponds to $j_1=1,j_2=0$, up until $j=P=(d_4+1)^2$, which corresponds to $j_1=d_4, j_2=d_4$. 
Then, \eqref{eq:nl_4states} can be written as
\begin{equation}
\dot{x}_4 = a_4 x_4 + \sum^P_{j=1} \sum^{d_3}_{j_3=0} \tilde{\alpha}^3_{j, j_3}\zeta_j x^{j_3}_3.
\end{equation}
Let $V_4=\{\zeta_1 x^{0}_3, \ldots, \zeta_P x^{d_4}_3\}$ and $W_4=\{x_4\} \cup V_4$, while $\Phi=\vc{\bigcup_{i=1}^4 W_i}$. By calculating $\dot{\Phi}$, we get the terms associated with $W_1, W_2, W_3$ as before and
\begin{align}
\frac{\dif}{\dif t}\left(\zeta_j x^{j_3}_3\right)&=\dot{\zeta}_j x^{j_3}_3+j_3\zeta_j x^{j_3-1}_3 \dot{x}_3 \label{eq:proof:3} \\
&=j_3 a_3 \underbrace{\zeta_j x^{j_3}_3}_a  + j_3\sum^{P}_{\tilde{j}=1}\tilde{\alpha}^2_{\tilde{j}}\underbrace{\zeta_{j+n_{\tilde{j}}} x^{j_3-1}_3}_{b}+ \underbrace{\dot{\zeta}_{j} x^{j_3}_3}_c \notag
\end{align}
\begin{itemize}
\item The terms $a$ are already contained in $V_4$.
\item For the terms $b$, we can observe that the power $j_3$ decreases by 1 and the powers of $x_1$ and $x_2$ within $\zeta$ increase by at most $d_3$ (which is finite), encoded in terms of $n_{\tilde{j}}$. Applying the same iterations as in case $n=3$, we can construct a $V_4$ such that $\mathfrak{D}_\mathrm{b}(V_4) \setminus V_4
\subseteq\{\zeta_1,\ldots,\zeta_{n_\zeta}\}$. We can observe that the $\zeta_j$ terms are in the form of the $b$ terms in case of $n=3$, hence the same procedure can be further applied till $\mathfrak{D}_\mathrm{b}(V_4) \setminus V_4
\subseteq\{x_1^{0},\ldots,x_1^{n_1}\}$.
\item
For the terms $c$,  $\frac{\dif}{\dif t}\zeta_{j}$ leads to a decrease of the orders of $x_1^{j_1}$ and $x_2^{j_2}$ in the terms $\zeta$.  Introduce the operator $\mathfrak{D}_\mathrm{c}$ such that $\mathfrak{D}_\mathrm{c}(\zeta_j x^{j_3}_3)=\{\dot{\zeta}_j x^{j_3}_3\}_{j=1}^{P}$, i.e., it gives the $c$ terms of \eqref{eq:proof:3}. Then let $V_4 \leftarrow V_4 \cup \mathfrak{D}_\mathrm{c}(V_4)$.  Repeating the process for a finite number of steps gives that $\mathfrak{D}_\mathrm{c}(V_4) \setminus V_4 \subseteq\{x_3^{0},\ldots,x_3^{n_3}\}$. Note that the empty set is also a subset and that the $b$ and the $c$ terms are iterated together.
\end{itemize}
 Hence, based on case $n=3$, we know that for $V_4 \leftarrow V_4 \cup \mathfrak{D}_\mathrm{c}(V_4)$ taking $W_4=\{x_4\} \cup V_4$ and $\Phi=\vc{\bigcup_{i=1}^4 W_i}$ will ensure that  condition \eqref{eq:conditions_b} holds true.
\item n+1 states ($n+1$ order system): \\
Assume that for $\Phi=\vc{\bigcup_{i=1}^n W_i}$, condition \eqref{eq:conditions_b} holds true in the $n^\mathrm{th}$-order case.
The dynamics of the $n+1$ order system is described by \eqref{eq:nl_aut}, together with:
\begin{equation}\label{eq:nl_np1states}
\dot{x}_{n+1}=a_n x_n+ \sum^{d_{n+1}}_{j_1=0}\dots \sum^{d_{n+1}}_{j_n=0}\alpha^{n+1}_{j_1 \dots j_n}x^{j_1}_1 \dots x^{j_n}_n.
\end{equation}
Similar to the $n=4$ case, introduce $\zeta_j = x^{j_1}_1\dots x^{j_{n-1}}_{n-1}$, with $j=1+\sum^{n-1}_{k=1}j_k(d_{n+1}+1)^{k-1}$ and $P=(d_{n+1}+1)^{n-1}$.
With this notation, \eqref{eq:nl_np1states} is equivalent to:
\begin{equation}
\dot{x}_{n+1} = a_{n+1} x_{n+1} + \sum^P_{j=1} \sum^{d_{n+1}}_{j_n=0} \tilde{\alpha}^{n+1}_{j, j_n}\zeta_j x^{j_n}_n.
\end{equation}
Let $V_{n+1}=\{\zeta_1 x^{0}_{n}, \ldots, \zeta_P x^{d_{n+1}}_n\}$ and $W_{n+1}=\{x_{n+1}\} \cup V_{n+1}$, while $\Phi=\vc{\bigcup_{i=1}^{n+1} W_{i}}$. By calculating $\dot{\Phi}$, we get the terms associated with $W_1, \ldots, W_{n}$ as before and
\begin{align}
\frac{\dif}{\dif t}\left(\zeta_j x^{j_n}_n\right)&=\dot{\zeta}_j x^{j_n}_n+j_n\zeta_j x^{j_n-1}_n \dot{x}_n \label{eq:proof:4} \\
&=j_n a_n \underbrace{\zeta_j x^{j_n}_n}_a  + j_n\sum^{P}_{\tilde{j}=1}\tilde{\alpha}^n_{\tilde{j}}\underbrace{\zeta_{j+n_{\tilde{j}}} x^{j_n-1}_n}_{b}+ \underbrace{\dot{\zeta}_{j}x_n^{j_n}}_c \notag
\end{align}
\begin{figure}[t!]
\begin{center}
\includegraphics[width=.4\textwidth]{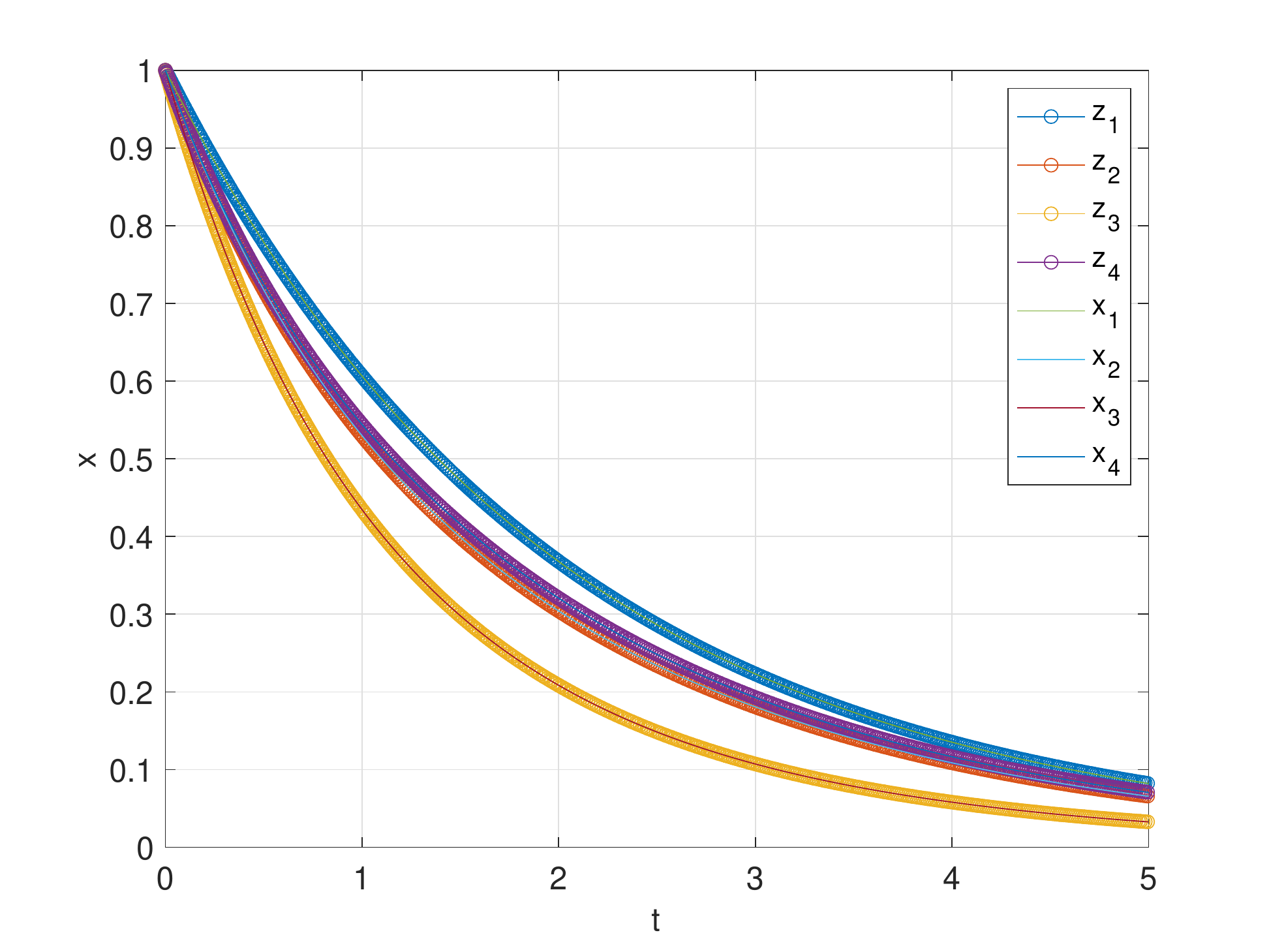}\vspace{-.4cm}    
\caption{State trajectories of the original nonlinear system representation \eqref{eq:ex_state1}-\eqref{eq:ex_state4} and the Koopman embedding \eqref{eq:koop_aut}.} 
\label{fig:autonomous_traj}
\end{center}
\end{figure}
\begin{itemize}
\item The terms $a$ are already contained in $W_n$.
\item We can observe that the power $j_n$ decreases by 1 and the powers of $x_i$ ($i\in\left\lbrace 1,\dots, n-1 \right\rbrace$) within $\zeta$ increase by at most $d_n$ (which is finite), encoded in terms of $n_{\tilde{j}}$. Applying the same iterations as in case $n=4$, recursively leads to $\mathfrak{D}_\mathrm{b}(V_{n+1}) \setminus V_{n+1}
\subseteq\{x_1^{0},\ldots,x_1^{n_1}\}$ in a finite number of steps.
\item As seen in $n=4$, taking $\frac{\dif}{\dif t}\zeta_{j}$ for the terms $c$, leads to a decrease of the orders of $x_1^{j_1},\ldots,x_{n}^{j_{n}}$ in the terms $\zeta$. By using $V_{n+1} \leftarrow V_{n+1} \cup \mathfrak{D}_\mathrm{c}(V_{n+1})$ in a finite number of steps leads to
$\mathfrak{D}_\mathrm{c}(V_{n+1}) \setminus V_{n+1} \subseteq\{x_{n}^{0},\ldots,x_{n}^{n_{n}}\}$. As noted before, the empty set is also a subset and the terms $b$ and $c$ are iterated together.
\end{itemize}
 Hence, based on case $n$, we know that for $V_{n+1} \leftarrow V_{n+1} \cup \mathfrak{D}_\mathrm{c}(V_{n+1})$ taking $W_{n+1}=\{x_{n+1}\} \cup V_{n+1}$ and $\Phi=\vc{\bigcup_{i=1}^{n+1} W_i}$ will ensure that  condition \eqref{eq:conditions_b} holds true. This completes the proof.
\end{itemize}
\par
This shows that for an autonomous polynomial nonlinear system with the dynamics described by \eqref{eq:nl_aut}, there exists a finite dimensional lifting $\Phi$, containing the states and polynomial terms, satisfying $\dot{\Phi} =\frac{\partial \Phi}{\partial x}f \in \mathrm{span}\{\Phi\}$. This implies that there exists a square real matrix $A$ such that $\dot{\Phi}(x)=A\Phi(x)$.
\end{pf}
\begin{figure}[t!]
\begin{center}
\includegraphics[width=.4\textwidth]{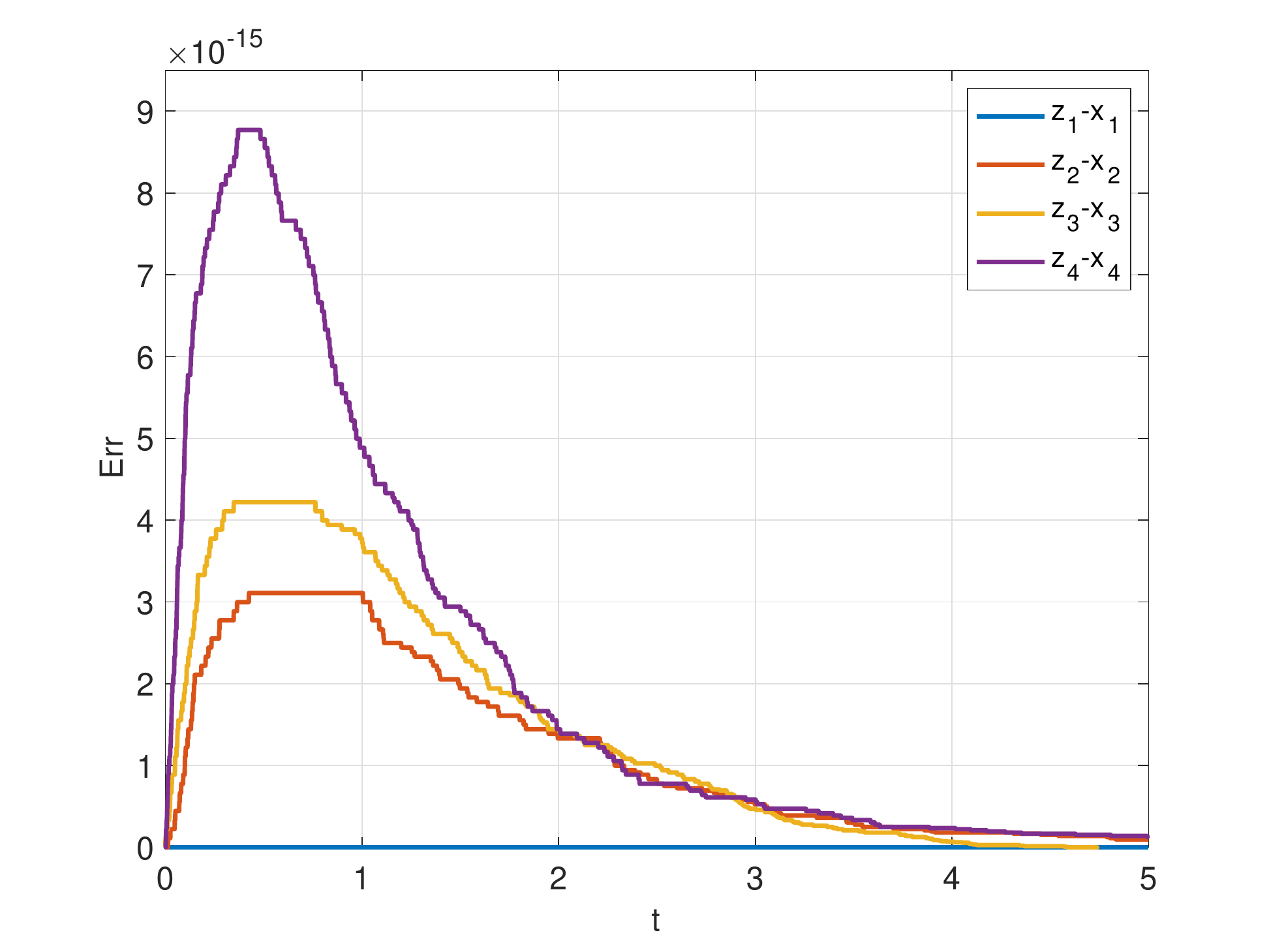}\vspace{-.4cm}    
\caption{Error between the state trajectories of the original nonlinear system representation \eqref{eq:ex_state1}-\eqref{eq:ex_state4} and the Koopman embedding \eqref{eq:koop_aut}.} 
\label{fig:autonomous_err}
\end{center}
\end{figure}
\subsection{Systems with input}\label{sec:sys_input}
Consider the following control affine nonlinear system:
\begin{equation}\label{eq:gen_input}
\dot{x}=f(x) +g(x)u,
\end{equation}
with the autonomous part given by \eqref{eq:nl_aut} and $g:\mathbb{R}^{n_{\mathrm{x}}}\rightarrow\mathbb{R}^{n_{\mathrm{x}}\times n_{\mathrm{u}}}$ and $u\in\mathbb{U}\subseteq \mathbb{R}^{n_{\mathrm{u}}}$. To obtain the lifted representation, one can use the sequential method described in \citep{Iacob:22}. First, an exact lifting of the autonomous part is assumed to exist, i.e. conditions \eqref{eq:conditions} hold. Next, the Koopman embedding is computed using the properites of the differential operator. Applying the lifting $\Phi$ and taking the time derivative, one obtains: 
\begin{equation}
\begin{split}
\dot{\Phi}&=\frac{\partial \Phi}{\partial x}(x)\dot{x}\\
&=\frac{\partial \Phi}{\partial x}(x)f(x) + \frac{\partial \Phi}{\partial x}(x)g(x) u.
\end{split}
\end{equation}
Using the equivalence of conditions \eqref{eq:conditions_a} and \eqref{eq:conditions_b}, an associated Koopman embedding of \eqref{eq:gen_input} is:
\begin{equation}\label{eq:koop_gen_input}
\dot{\Phi}(x)=A\Phi(x) + B(x) u,
\end{equation}
with $B(x)=\frac{\partial \Phi}{\partial x}(x)g(x)$. As described in \citep{Iacob:22}, one can further express \eqref{eq:koop_gen_input} as a \textit{linear parameter varying} (LPV) Koopman representation by introducing a scheduling map $p=\mu(z)$, where $z=\Phi(x)$ and defining $B_{\mathrm{z}}\circ z = B$. Then, the LPV Koopman model is described by:
\begin{equation}\label{eq:koop_LPV}
\dot{z}=Az + B_{\mathrm{z}}(p)u,
\end{equation}
with $z(0)=\Phi(x(0))$.
\section{Example}\label{sec:example}
This section presents the embedding of an example $4$-dimensional system and shows simulation results for both autonomous and input-driven operation.
\subsection{Autonomous case}
Consider the following $4^{\mathrm{th}}$ order system:
\begin{align}
\dot{x}_1 &= a_1 x_1 \label{eq:ex_state1}\\
\dot{x}_2 &= a_2 x_2 + \alpha^2_3 x^3_1 \label{eq:ex_state2}\\
\dot{x}_3 &= a_3 x_3 + \alpha^3_{11} x_1x_2 + \alpha^3_{02} x^2_2 \label{eq:ex_state3}\\
\dot{x}_4 &= a_4 x_4 + \alpha^4_{111} x_1x_2x_3. \label{eq:ex_state4}
\end{align}
We can apply the procedure discussed in Section \ref{sec:embedding} per state equation to find the observable functions. The resulting lifting functions are as follows: $W_1=\left\lbrace x_1
\right\rbrace$, $W_2=\left\lbrace x_2,x^3_1
\right\rbrace$,  $W_3=\left\lbrace x_3,x_1x_2,x^2_2,x^4_1,x^3_1x_2,x^6_1
\right\rbrace$, and $W_4=\{ x_4,x_1x_2x_3,x^4_1x_3,x^2_1x_2,x_1x^3_2,x^5_1x_2,x^4_1x_2,x^8_1,x^7_1x_2,x^{10}_1
\}$. \\
Then, the entire lifting set is $\Phi = \vc{W_1,W_2,W_3,W_4}$. For easier interpretability, we can write the observables such that: $\Phi(x) = [ x_1 \ x_2 \ x_3 \ x_4 \ \bar{\Phi}_1^\top \ \bar{\Phi}_2^\top \ \bar{\Phi}_3^\top \ \bar{\Phi}_4^\top ]^\top$ and $\bar{\Phi}_i$ contains the elements of $W_i$, in order, without the state $x_i$. Performing the derivations as described in the proof, we obtain a finite dimensional Koopman representation of the form:
\begin{equation}\label{eq:koop_aut}
\begin{split}
\dot{z}&=Az\\
x&=Cz,
\end{split}
\end{equation}
with $z(t)=\Phi(x(t))$, $A\in\mathbb{R}^{19\times 19}$ and $C=[I_4\; 0_{4\times 15}]$. The structure of the state matrix $A$ is detailed in the Appendix. To compare the obtained Koopman representation and the original system description, consider $a_1=a_2=a_3=a_4 = -0.5$, $\alpha^2_3=\alpha^3_{11}=\alpha^3_{02}=\alpha^4_{111}=-0.2$ and $x_0 = [1\;1\;1\;1]^\top$. We can obtain solution trajectories of these two representations by a Runge-Kutta $4^\mathrm{th}$ order solver. Furthermore, once the initial condition is lifted, i.e. $z(0)=\Phi(x(0))$, the dynamics of the Koopman model are driven forward linearly, as described by \eqref{eq:koop_aut}. The simulation results and solution trajectories are depicted in Fig \ref{fig:autonomous_traj}. As it can be observed, there is an exact overlap between the state trajectories of the original system description and the state trajectories obtained from the lifted model ($z_{1\rightarrow 4}$ correspond to $x_{1\rightarrow 4}$). Fig. \ref{fig:autonomous_err} shows that the obtained error is in the order of magnitude of $10^{-15}$, which can be attributed to numerical artifacts.
\begin{figure}[t!]
\begin{center}
\includegraphics[width=.45\textwidth]{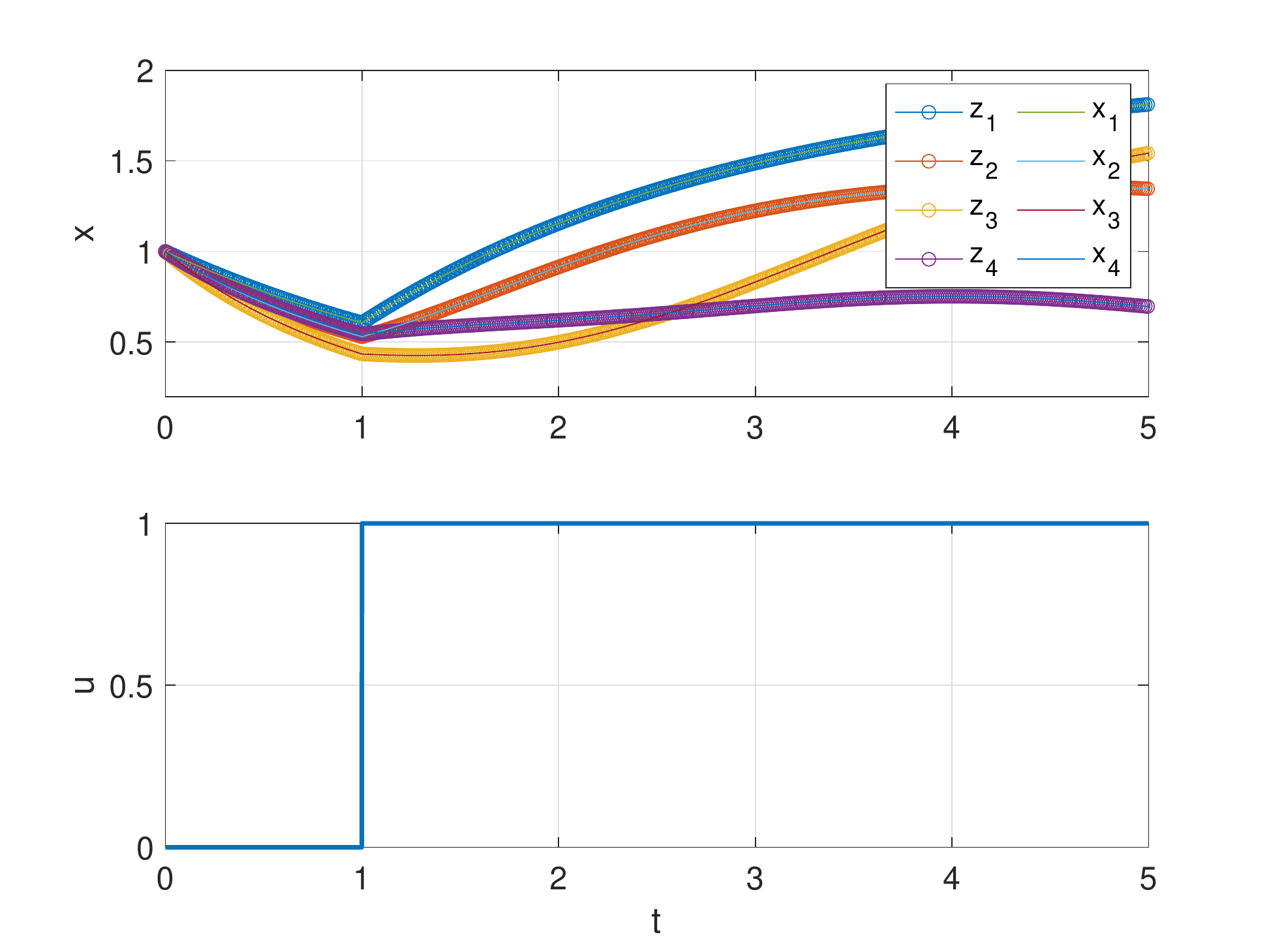}\vspace{-.4cm}    
\caption{State trajectories of the original nonlinear system detailed in Section \ref{sec:example_input} and the Koopman embedding \eqref{eq:koop_input}.} 
\label{fig:input_traj}
\end{center}
\end{figure}
\subsection{Input-driven case}\label{sec:example_input}
\begin{figure}[t!]
\begin{center}
\includegraphics[width=.45\textwidth]{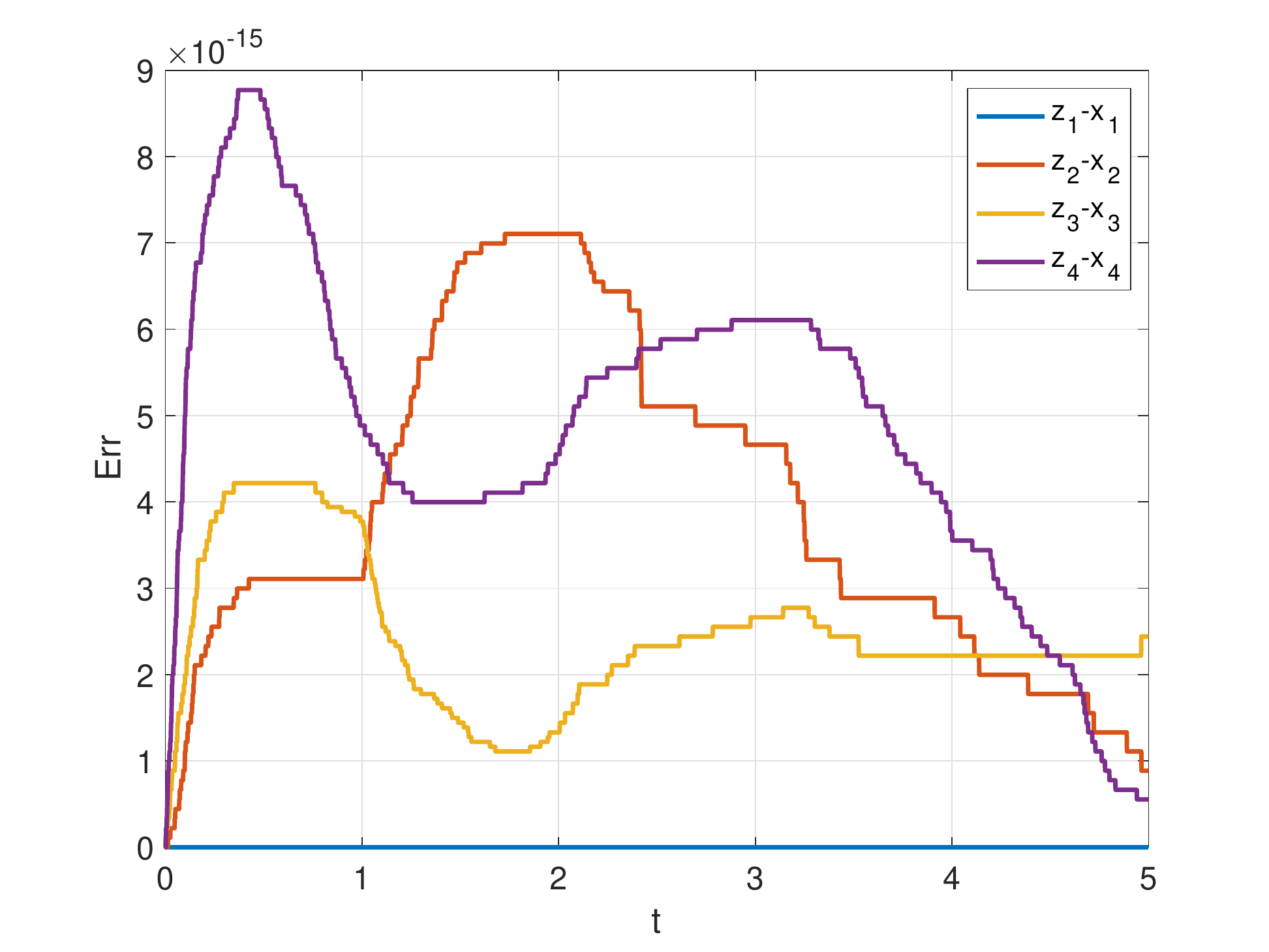}\vspace{-.4cm}    
\caption{Error between the state trajectories of the original nonlinear system detailed in Section \ref{sec:example_input} and the Koopman embedding \eqref{eq:koop_input} under a step input.} 
\label{fig:input_err}
\end{center}
\end{figure}

Consider a control affine nonlinear system \eqref{eq:gen_input}, with the autonomous part given by the equations \eqref{eq:ex_state1}-\eqref{eq:ex_state4} and $g(x) = \left[1\; x_1\;x^2_2\; \sin (x_3) \right]^\top$. Applying the lifting procedure described in Section \ref{sec:sys_input}, we can derive an exact LPV Koopman model:
\begin{equation}\label{eq:koop_input}
\begin{split}
\dot{z}&=Az + B_{\mathrm{z}}(p)u\\
x&=Cz,
\end{split}
\end{equation}
with $C=[I_4\; 0_{4\times 15}]$, $z(t)=\Phi(x(t))$ and $p = z$. Note that the state matrix $A$ coincides with the autonomous case. The explicit form of $B(x)$ (and, in turn, $B_{\mathrm{z}}$) is omitted due to space constraints, but it can be easily computed by multiplying $\frac{\partial \Phi}{\partial x}$ with $g(x)$. The structure of $\frac{\partial \Phi}{\partial x}(x)$ is given in the Appendix. We use the same coefficient values as in the autonomous case and consider a step input. After lifting the initial state $z(0)=\Phi(x(0))$, the dynamics of the Koopman representation are simulated forward in time by \eqref{eq:koop_input}.  Fig. \ref{fig:input_traj} shows the solution trajectories of both the original and the lifted system representations. As in the autonomous case, there is an exact overlap, with the error between the state trajectories being in the order of magnitude of $10^{-15}$, only due to numerical integration errors. This is depicted in Fig. \ref{fig:input_err}.
\section{Conclusion}\label{sec:conclusion}
The present paper shows that a finite, exact Koopman embedding exists for a specific system class and an approach is provided to obtain this embedding. Furthermore, as shown, the step to embed nonlinear systems with input is easily achieved once the autonomous part is lifted. Future work will focus on extending the current system description to a more general class of nonlinear systems. 
\balance
\bibliography{ifacconf}

\onecolumn
\appendix
\section{Matrices}
\setcounter{MaxMatrixCols}{19}
\resizebox{\textwidth}{!}{
$A = \begin{bmatrix}
a_1 & 0 & 0 & 0 & 0 & 0 & 0 & 0 & 0 & 0 & 0 & 0 & 0 & 0 & 0 & 0 & 0 & 0 & 0\\
0 & a_2 & 0 & 0 & \alpha^2_3 & 0 & 0 & 0 & 0 & 0 & 0 & 0 & 0 & 0 & 0 & 0 & 0 & 0 & 0 \\
0 & 0 & a_3 & 0 & 0 & \alpha^3_{11} & \alpha^3_{02} & 0 & 0 & 0 & 0 & 0 & 0 & 0 & 0 & 0 & 0 & 0 & 0 \\
0 & 0 & 0 & a_4 & 0 & 0 & 0 & 0 & 0 & 0 & \alpha^4_{111} & 0 & 0 & 0 & 0 & 0 & 0 & 0 & 0 \\
0 & 0 & 0 & 0 & 3a_1 & 0 & 0 & 0 & 0 & 0 & 0 & 0 & 0 & 0 & 0 & 0 & 0 & 0 & 0 \\
0 & 0 & 0 & 0 & 0 & a_1+a_2 & 0 & \alpha^2_3 & 0 & 0 & 0 & 0 & 0 & 0 & 0 & 0 & 0 & 0 & 0 \\
0 & 0 & 0 & 0 & 0 & 0 & 2a_2 & 0 & 2\alpha^2_3 & 0 & 0 & 0 & 0 & 0 & 0 & 0 & 0 & 0 & 0 \\
0 & 0 & 0 & 0 & 0 & 0 & 0 & 4a_1 & 0 & 0 & 0 & 0 & 0 & 0 & 0 & 0 & 0 & 0 & 0 \\
0 & 0 & 0 & 0 & 0 & 0 & 0 & 0 & 3a_1+a_2 & \alpha^2_3 & 0 & 0 & 0 & 0 & 0 & 0 & 0 & 0 & 0 \\
0 & 0 & 0 & 0 & 0 & 0 & 0 & 0 & 0 & 6a_1 & 0 & 0 & 0 & 0 & 0 & 0 & 0 & 0 & 0 \\
0 & 0 & 0 & 0 & 0 & 0 & 0 & 0 & 0 & 0 & a_1+a_2+a_3 & \alpha^2_3 & \alpha^3_{11} & \alpha^3_{02} & 0 & 0 & 0 & 0 & 0\\
0 & 0 & 0 & 0 & 0 & 0 & 0 & 0 & 0 & 0 & 0 & 4a_1+a_3 & 0 & 0 & \alpha^3_{11} & \alpha^3_{02} & 0 & 0 & 0\\
0 & 0 & 0 & 0 & 0 & 0 & 0 & 0 & 0 & 0 & 0 & 0 & 2a_1+2a_2 & 0 & 2\alpha^2_3 & 0 & 0 & 0 & 0\\
0 & 0 & 0 & 0 & 0 & 0 & 0 & 0 & 0 & 0 & 0 & 0 & 0 & a_1+3a_2 & 0 & 3\alpha^2_3 & 0 & 0 & 0\\
0 & 0 & 0 & 0 & 0 & 0 & 0 & 0 & 0 & 0 & 0 & 0 & 0 & 0 & 5a_1+a_2 & 0 & \alpha^2_3 & 0 & 0\\
0 & 0 & 0 & 0 & 0 & 0 & 0 & 0 & 0 & 0 & 0 & 0 & 0 & 0 & 0 & 4a_1+2a_2 & 0 & 2\alpha^2_3 & 0\\
0 & 0 & 0 & 0 & 0 & 0 & 0 & 0 & 0 & 0 & 0 & 0 & 0 & 0 & 0 & 0 & 8a_1 & 0 & 0\\
0 & 0 & 0 & 0 & 0 & 0 & 0 & 0 & 0 & 0 & 0 & 0 & 0 & 0 & 0 & 0 & 0 & 7a_1+a_2 & \alpha^2_3\\
0 & 0 & 0 & 0 & 0 & 0 & 0 & 0 & 0 & 0 & 0 & 0 & 0 & 0 & 0 & 0 & 0 & 0 & 10a_1
\end{bmatrix}$
}
\par
\setcounter{MaxMatrixCols}{19}
\resizebox{\textwidth}{!}{
$\frac{\partial \Phi}{\partial x}(x) = \begin{bmatrix}
1 & 0 & 0 & 0 & 3x^2_1 & x_2 & 0 & 4x_1^3 & 3x_1^2x_2 & 6x_1^5 & x_2x_3 & 4x_1^3x_3 & 2x_1x_2^2 & x_2^3 & 5x_1^4x_2 & 4x_1^3x^2_2 & 8x_1^7 & 7x_1^6x_2 & 10 x_1^9\\
0 & 1 & 0 & 0 & 0 & x_1 & 2x_2 & 0 & x_1^3 & 0 & x_1x_3 & 0 & 2x_1^2x_2 & 3x_1x_2^2 & x_1^5 & 2x_1^4 x_2 & 0 & x_1^7 & 0\\
0 & 0 & 1 & 0 & 0 & 0 & 0 & 0 & 0 & 0 & x_1x_2 & x_1^4 & 0 & 0 & 0 & 0 & 0 & 0 & 0\\
0 & 0 & 0 & 1 &  0 & 0 & 0 & 0 & 0 & 0 & 0 & 0 & 0 & 0 & 0 & 0 & 0 & 0 & 0
\end{bmatrix}^\top
$
}

                                                   







\end{document}